\documentstyle[twocolumn,prl,aps]{revtex}

\begin{document}
\draft
\preprint{Dortmund, May 1995}

\title{Theory of Non-Reciprocal Optical Effects in 
Antiferromagnets:
The Case Cr$_2$O$_3$}
\author{V. N. Muthukumar, Roser Valent\'\i\ and Claudius Gros
       }
\address{Institut f\"ur Physik, Universit\"at Dortmund,
         44221 Dortmund, Germany
        }
\date{\today}
\maketitle
\begin{abstract}

A microscopic model of non-reciprocal optical effects in
antiferromagnets is developed by considering the case of Cr$_2$O$_3$
where such effects have been observed. These effects are due to a
{\em direct} coupling between light and the antiferromagnetic order 
parameter. This coupling is mediated by the spin-orbit interaction 
and involves an interplay between the breaking of inversion symmetry due 
to the 
antiferromagnetic order parameter and the trigonal field contribution 
to the ligand field at the magnetic ion. We evaluate
the matrix elements relevant for the non-reciprocal second harmonic 
generation and gyrotropic birefringence. 

\end{abstract}
\pacs{42.65.-k,78.20.-e,78.20.Ls}
\section{Introduction}
The study of optical phenomena in magnetic substances has always been an
interesting area of research. Optical effects exhibited by magnetic
substances can be classified broadly into two categories, reciprocal and
non-reciprocal \cite{S_P_92}. Reciprocal optical phenomena are those that
cannot distinguish between magnetic states that are related to one
another by time reversal. Typical examples of such phenomena are those
that involve scattering of light by magnetic excitations. Non-reciprocal
phenomena, on the other hand, can distinguish between two magnetic
states that are related to each other by time reversal. A classic
example of such a phenomenon is the Faraday effect discovered by Faraday
in the last century. The Faraday effect in ferromagnets, for instance,
appears as a rotation of the plane of polarization of light which is
incident along the axis of magnetization. This rotation is
non-reciprocal in the sense that it changes sign on reversing the
direction of magnetization (which is equivalent to time reversal). Such
an effect can be understood in terms of an interaction between the
internal molecular field of the ferromagnet and the incident
electromagnetic radiation mediated by the spin-orbit coupling 
\cite{Argyres}. Clearly, such an effect would not only be exhibited by 
ferromagnets but by any substance that has a net (non-zero) magnetic 
moment such as ferrimagnets or paramagnets where a magnetic moment is 
induced by the application of an external field.

It is then but natural to ask what happens in the case of
antiferromagnets. At first sight, it would seem that it is impossible
for light to distinguish between states that are related by time
reversal as, in this case, there is no net magnetic moment (i.e., the
total molecular field is zero unlike in ferromagnets). However, this is
not true. In fact, it has been known for a long time from symmetry 
arguments that certain classes of antiferromagnets can show a variety of
non-reciprocal phenomena \cite{Dzyalo,Brown_63}, although till recently,
there have been no
reports of experimental observations of such phenomena. The utility of 
such experiments, where possible, can hardly be overemphasized since 
experiments such as Raman scattering etc., only probe the antiferromagnetic 
structure {\em indirectly} by coupling to the magnetic excitations 
rather than the magnetic ordering itself \cite{Raman}.

In this context therefore, the discovery of non-reciprocal optical effects 
below the N\'eel temperature, $T_N$, in optical experiments on Cr$_2$O$_3$ 
\cite{Krichevtsov_93,Fiebig_94} comes as a breakthrough in the 
study of antiferromagnetic ordering by light. Krichevtsov {\it et al.} 
\cite{Krichevtsov_93} reported the first experimental observation of 
spontaneous non-reciprocal rotation and circular dichroism below the 
N\'eel temperature of Cr$_2$O$_3$.  Fiebig {\it et al.} \cite{Fiebig_94} 
found that antiferromagnetic domains could be observed {\it directly} 
by non-reciprocal second harmonic generation (SHG), leading to the 
first photographs ever of antiferromagnetic domains \cite{APL,note_Roth}. 
These experiments show that light can indeed couple directly to the 
antiferromagnetic order parameter, thereby leading to non-reciprocal effects. 
As mentioned earlier, though such a coupling was anticipated from 
symmetry considerations, no microscopic mechanism has been presented so far. 

In this paper, we present, in detail, a microscopic mechanism that
describes all non-reciprocal optical effects by considering the case of
Cr$_2$O$_3$ where such effects have been observed experimentally. In
an earlier paper \cite{short_pap}, we showed
that all non-reciprocal effects in Cr$_2$O$_3$ 
can be explained by the fact that {\em electric
dipole} transitions are allowed in Cr$_2$O$_3$ below the N\'eel
temperature. Here we highlight the actual evaluation of the
matrix elements relevant for the non-reciprocal non-linear
susceptibilities. 
The evaluation of the matrix elements is performed within 
a new cluster model for Cr$_2$O$_3$, which contains the full crystal 
symmetry of Cr$_2$O$_3$ and which allows the
orders of magnitude of all matrix elements contributing 
to the non-reciprocal phenomena in Cr$_2$O$_3$ to be predicted.
We apply the microscopic model to the observed phenomenon of SHG
\cite{Fiebig_94} and explain how antiferromagnetic domains can be
distinguished experimentally. We also apply our model to another
non-reciprocal effect seen experimentally in Cr$_2$O$_3$ viz., 
gyrotropic birefringence \cite{Krichevtsov_93} and solve the long-standing 
question regarding its magnitude. Our model can also be used to
describe non-reciprocal optical effects in other antiferromagnets where
inversion symmetry is broken below $T_N$.

The paper is organized as follows. In section II, we illustrate how
symmetry arguments can be used to study non-reciprocal effects by
considering the macroscopic theory of SHG in Cr$_2$O$_3$. In section
III, we present the microscopic theory of SHG which allows us to obtain all
polarization selection rules and magnitudes of the nonlinear
susceptibilities. In section IV, we use our microscopic model to explain
the phenomenon of gyrotropic birefringence and the associated
magnetoelectric effect. A summary of our results and conclusions are
presented in section V.

\section{Macroscopic Theory}
A clue to the origin of non-reciprocal effects can be obtained from
macroscopic (symmetry) considerations of the susceptibility tensors. 
The optical properties of a medium are characterized by linear and
nonlinear susceptibility tensors. If the susceptibility tensors are 
known for a given medium up to a certain order $n$, then, at least in 
principle, the $n$th-order nonlinear optical effects in the medium can 
be predicted from Maxwell's equations.
Physically, the susceptibility is related to the microscopic structure of the
medium and can be properly evaluated only by doing a 
 full quantum-mechanical calculation as we will show in the next section.
Nevertheless, one can get some information about the susceptibilities just
from symmetry considerations as dictated by Neumann's principle,which  
states \cite{Birss} 
that any symmetry which is exhibited by the point group of the 
crystal is possessed by every physical property of the crystal. Then, 
to investigate the effect of crystal symmetry on the components of the 
susceptibility tensors, it is necessary to enforce 
the requirement that the tensor be invariant under all the permissible 
symmetry operators appropriate to the particular crystal class.  
Accordingly, some tensor elements are zero or are related to 
others, thereby reducing the total number of independent tensor elements.
The application of Neumann's principle to four-dimensional 
space-time remains valid only for static properties. For dynamic processes, 
like non-reciprocal SHG, where there is a preferred direction of time, 
one considers only those symmetry operations that do not include time 
reversal in classifying the allowed tensors, {\it i.e.}, only the
spatial symmetry is used to simplify the form of the susceptibility
tensors. Though one can still classify the tensors as $i$ and
$c$ (time-symmetric/antisymmetric) tensors, one cannot use this classification
to obtain those that are allowed.
To illustrate these notions, let us now consider the macroscopic theory
of SHG in Cr$_2$O$_3$ \cite{Fiebig_94}.  We shall consider in detail the
case where the spins are oriented parallel to the 
crystallographic $z$ axis and laser
light propagates along the $z$ direction and then we shall generalize to the
case of arbitrary direction for the orientation of the spins and for the
propagation of the laser light.

\subsection{Spins and laser light parallel to $z$ axis}

Above the N\'eel temperature, $T_N \simeq 307K$, Cr$_2$O$_3$
 crystallizes in the centrosymmetric 
point group $D_{3d}$. The four Cr$^{3+}$ ions in the unit cell occupy 
equivalent $c$ positions along the $3_z$ (optic) axis. 
Since this structure has a center of inversion, parity considerations forbid
electric dipole transitions in SHG. Magnetic dipole 
transitions that are related to the existence of an axial tensor of 
odd rank, or electric quadrupole transitions related to the existence of
a polar tensor of even rank are however allowed.
Below $T_N$, the spins of the Cr$^{3+}$ ions orient along the $z$-axis. 
The spin ordering breaks time
reversal symmetry, R, and as  SHG is a dynamic process, 
only symmetry operations of the crystal that do not include R may be 
used to classify the allowed tensors for the susceptibilities. For 
Cr$_2$O$_3$, the remaining invariant subgroup is $D_3$. New tensors 
are allowed in this point group, for instance, a polar tensor of odd rank, 
that allow electric dipole transitions in SHG.

From Maxwell's equations, the source term corresponding to SHG can be derived 
by considering the contributions to (nonlinear) magnetization,
${\bf M}^{(2 \omega)}= \epsilon_o c ~
{\bf{\gamma}}^{(2 \omega)} : {\bf E}^{(\omega)}~ {\bf E}^{(\omega)}$ and 
polarization, 
${\bf P}^{(2 \omega)}= \epsilon_o {\bf \chi}^{(2 \omega)} :{\bf E}^{(\omega)}~ 
{\bf E}^{(\omega)}$, in the point group $D_3$. In general, one should
also consider contributions to nonlinear susceptibilities related to
electric quadrupole transitions, {\it viz.}, 
${\bf Q}^{(2\omega)} =
\frac{-ic \epsilon_o}{2 \omega} \tilde{{\bf{\chi}}}^{(2\omega)} :
{\bf E}^{(\omega)}~{\bf E}^{(\omega)}$.
Since the basic physics associated with such transitions is not very
different from the magnetic dipole transitions, we will in the following
concentrate on the electric and magnetic dipole contributions to the 
nonlinear susceptibilities.
If we assume that laser light propagates along the optic axis, there is 
only one independent component $\gamma_m$ $\equiv$ $\gamma_{yyy}$ $=$ 
$-\gamma_{yxx}$ $=$ $-\gamma_{xyx}$ $=$ $-\gamma_{xxy}$ of the nonlinear 
magnetic susceptibility (axial tensor of third rank) and analogously 
for the nonlinear electric susceptibility (polar tensor of third rank), 
$\chi_e$ $\equiv$ $\chi_{yyy}$ $=$ $-\chi_{yxx}$ $=$
$-\chi_{xyx}$ $=$ $-\chi_{xxy}$.
The source term ${\bf S}({\bf r},t)$ in the wave equation,
\begin{equation}
 [ {\bf{\nabla}} \times ( {\bf{\nabla}} \times ) + \left( 1/c^2\right)
\partial^2/\partial t^2 ] {\bf{E}}( {\bf r}, t) = 
-{\bf{S}}({\bf r}, t),
\end{equation}
 can be written in a dipole expansion as \cite{Rosen},
\begin{equation}
\begin{array}{rcl}
{\bf{S}}({\bf r},t) &=& 
\mu_o~ \left( \frac{{\partial}^2 {\bf{P}}({\bf r},t)}{\partial t^2} +
{\bf{\nabla}} \times \frac{\partial {\bf{M}}({\bf r},t)}{\partial t}
        \,\right.\\
&& \qquad\left.-\,
  {\bf \nabla} \frac{\partial^2
 {\bf Q} ({\bf r},t)}{\partial t^2} + \dots\,
\right)~~.
\end{array}
\label{source}   
\end{equation}
If we now assume that {\bf E}, {\bf P} and {\bf M} can be decomposed 
in a set of plane waves and consider a circular basis with 
${\bf{E}} = E_+ {\hat{e}}_++ E_- {\hat{e}}_- + E_z {\hat{e}}_z $, where 
${\hat{e}}_{\pm} = \frac{1}{\sqrt{2}} ({\hat{e}}_x {\pm} i {\hat{e}}_y)$, 
one obtains (note the different basis choice with respect
 to refs. \cite{Fiebig_94} and \cite{short_pap}), 
\begin{equation}
{\bf{S}} = \left( \begin{array}{c}
S_+ \\ S_- \\ S_z \end{array} \right) = \frac{4 \sqrt{2} {\omega}^2} {c^2}  
\left(\begin{array}{c}
(\gamma_m - i \chi_e) E_-^2 \\  (\gamma_m + i \chi_e) E_+^2  \\ 0
\end{array} \right) ~~,  
\label{S}
\end{equation}      
where $\hbar \omega$ is the energy of the incoming light beam. 
 Note that incoming right circularly polarized light ($E_+$)
 leads to left circularly
polarized light ($E_-$) and vice versa in SHG.
Above $T_N$, the electric dipole contributions disappear 
($\chi_e \equiv 0$) and therefore the SHG intensities $I_{\pm}$ 
are identical while below $T_N$,
($\chi_e \neq 0$)  and the intensities 
$I_{\pm} \propto | \gamma_m \pm i \chi_e |^2 E_{\mp}^4$ 
are different for right and left
circularly polarized light, as observed experimentally 
\cite{Fiebig_94}. As the tensor $\chi$ appears only below $T_N$, it is
natural to assume that it is proportional to $\triangle (T)$, the
antiferromagnetic order parameter. Non-reciprocal SHG can then be
understood as arising from an interference between $\gamma_m$ and
$\chi_e$.

\subsection{Spins and laser light in arbitrary directions}

 An extended study of SHG in Cr$_2$O$_3$ can be carried out by
considering the propagation of light and the orientation of the
spins in the crystal to be arbitrary, the motivation of such a study
being the fact that application of an external magnetic field causes
 the spins to orient along different directions. 
The spin-flop phase which occurs in Cr$_2$O$_3$ when a static magnetic 
field is applied along the $z$ direction can be studied using these
results. In Table I, we present the results of the macroscopic
symmetry analysis below $T_N$ for light propagating 
along the $z$, $y$ and $x$ directions and for spins oriented
parallel to $z$, $y$ and $x$ axis.
The three components of the source term are indicated 
for the various cases.
Note that, in the first row which corresponds to the case that
the spins are aligned parallel to the $z$ axis, the interference
effect occuring for light
propagating along the $z$ direction is absent
when light propagates along the $y$ and $x$  
directions, as has been verified experimentally \cite{note_1}.

When the spins are aligned along the $z$, $y$ and
$x$ directions the invariant point subgroups of crystal symmetries are
$D_3$, $C_{1h}$ and $C_2$ respectively. Table I contains the
results for the source term components for all three possibilities.
Table II contains the components of the source term
corresponding to electric
quadrupole transitions for light propagating along the $z$, $y$
and $x$ directions and spins aligned parallel to the $z$ axis.


\section{Microscopic Theory of Second Harmonic Generation}
The macroscopic theory discussed so far is based purely on symmetry
considerations. It can neither provide magnitudes of the components of 
the magnetic ($\gamma$) and electric ($\chi$) susceptibilities
nor can it specify the details of the interaction and the origin of the
coupling between light and the antiferromagnetic order parameter.
Both these can be obtained only from a microscopic approach which is
described below. In what follows, we consider in detail the case
when the spins are aligned parallel to the $z$ axis and light
propagates along the same direction. This is the most representative
case for the observation of non-reciprocal phenomena in Cr$_2$O$_3$.
Elsewhere, we have shown that non-reciprocal effects can
be understood by the fact that electric dipole transitions are allowed
below $T_N$ \cite{short_pap}. We use this in calculating the nonlinear
susceptibilities of Cr$_2$O$_3$. The method of calculating nonlinear
susceptibilities of dielectric media was developed by Armstrong and
coworkers \cite{Armstrong_62}. In this approach, the current density
induced in the system by incident electromagnetic radiation is
calculated using semi classical perturbation theory. 

Let us consider the electromagnetic field to be a superposition of three
harmonic waves whose frequencies $\omega_i$, $i=1,2,3$, satisfy the condition
$\omega_3=\omega_1+\omega_2$. We choose the vector potential of this
field to be
$$
{\bf A}({\bf r},t) = \sum_i \hat{a}_i q_i \exp i({\bf k}\cdot {\bf
r}-\omega_it) +{\rm h.c.}~~.
$$
where $\hat{a}_i$ is a unit polarization vector and $q_i$ determines the
strength of the electric field, $E_i = q_i\omega_i/2$ (where the Coulomb
gauge is being assumed). The perturbation is then described by the usual
form of the interaction hamiltonian
$$
H_{{\rm int}}  =   \sum_j {-e \over m_e} {\bf A}_j \cdot {\bf p}_j 
+ {e^2 \over 2m_e} {\bf A}_j\cdot {\bf A}_j ~~.
$$
Assuming for simplicity that the ground state wave function $\Phi$
is a product of one-electron wave functions, $\Phi= \Pi_n \phi_n$, the
induced current density is given in the interaction representation
by
$$
\sum_n \langle \phi_n|e{\bf v}|\phi_n\rangle = 
\sum_n \langle \phi_n| {\bf p}-e{\bf A}|\phi_n\rangle{e\over m_e} ~~.
$$
Nonlinear susceptibilities are now calculated to the desired order in
field strengths. We are in particular interested in terms that are
quadratic in fields and the source term at $\omega_3=\omega_1+\omega_2$.
Thus we only consider those terms in the perturbation expansion for the
current density that are proportional to $q_1q_2$, {\it viz.},
proportional to $E_1E_2$ and with time dependence $\exp (\pm
i\omega_3t)$ and write
\[
\begin{array}{rcl}
\langle e{\bf v}(\omega_3)\rangle &=& 
\langle e{\bf v}^{(+)}(\omega_3) \exp (i\omega_3t)\rangle 
\\ &+& 
\langle e{\bf v}^{(-)}(\omega_3) \exp (-i\omega_3t)\rangle ~~.
\end{array}
\]
By multiplying the terms proportional to $\exp (\pm i\omega_3t)$ by
$\exp (\mp i{\bf k_3}\cdot {\bf r})$ and then taking the expectation
value, we get the induced current density corresponding to the source
term at $\omega_3$ and at ${\bf k}_1+{\bf k}_2 ={\bf k}_3$ as
\begin{equation}
\langle {\bf J}({\bf k}_3,\omega_3) \rangle  = \langle\Phi |e{\bf
v}^{(-)}(\omega_3) \exp i({\bf k}_3\cdot {\bf
r}-\omega_3t)|\Phi \rangle ~~.
\label{current}
\end{equation}
One can now obtain the expression for the nonlinear susceptibilities by
using 
the relation ${\bf J} = \mu_o \partial {\bf P}/\partial t$ as also
the defining relations for the susceptibilities that relate, for
example, ${\bf P}$ and ${\bf E}$.

We now turn our attention to the specific case of Cr$_2$O$_3$. As
explained in the previous section, we are interested in the nonlinear
polarization ${\bf P}^{(2\omega)}$ and the magnetization ${\bf
M}^{(2\omega)}$. The nonlinear susceptibility, $\chi$ is only allowed
below the N\'eel temperature where spatial inversion symmetry of the
crystal is broken by antiferromagnetic ordering. In an earlier paper
\cite{short_pap}, we have shown that this leads to electric dipole
transitions between the $^{4}A_2$ and $^{4}T_2$ levels of Cr$_2$O$_3$. 
Thus, below $T_N$, one can use the electric dipole approximation with
the interaction hamiltonian of the form 
$$
H = -e\sum_i {\bf E}_i \cdot {\bf r} ~~.
$$
A simple calculation now yields the result for the induced current,
\begin{eqnarray}
\langle {\bf J}({\bf k}_3,\omega_3) \rangle = {e^2 \over 2\hbar^2}
\sum_{|m\rangle,|n\rangle} \times
\qquad\qquad\qquad\qquad\qquad\nonumber \\
\Big[\,
{\langle \Phi |{\bf J}({\bf k}_3,\omega_3)|m\rangle
\langle m|E_{\mu}r_{\mu}|n\rangle 
\langle n|E_{\nu}r_{\nu}| \Phi\rangle
\over
(\omega_m-2\omega)(\omega_n-\omega)
} \nonumber \\
 + {\langle \Phi |E_{\mu}r_{\mu}|m\rangle 
\langle m|{\bf J}({\bf k}_3,\omega_3)|n\rangle
\langle n|E_{\nu}r_{\nu}| \Phi \rangle 
\over
(\omega_m+\omega)(\omega_n-\omega)}
\nonumber \\
 +{\langle \Phi |E_{\mu}r_{\mu}|m\rangle
\langle m|E_{\nu}r_{\nu}|n\rangle 
\langle n|{\bf J}({\bf k}_3,\omega_3)| \Phi \rangle 
\over
(\omega_n+2\omega)(\omega_m+\omega)}
\,\Big] ~~,
\label{jed}
\end{eqnarray}
where $E_{\mu}$ are the components of the incident electric fields
and the definition (\ref{current}) is to be used.
In the above expression we have also taken (as in SHG)
$\omega_1=\omega_2=\omega$ and $\omega_3 = 2\omega$. 
All factors $\exp(i{\bf k}\cdot {\bf
r})$ have been set to unity (electric dipole approximation). The problem
now reduces to that of evaluating the appropriate transition matrix
elements in the above expression   with the Cr$^{3+}$ wave functions. To
do this, we consider a cluster model that has the
correct symmetry of Cr$_2$O$_3$, {\it viz.}, $D_{3d}$. This model
\cite{short_pap}
contains only two Cr ions in a magnetic unit cell whereas in the
actual structure of Cr$_2$O$_3$  there are four. While this 
would lead to quantitative differences, the basic physics remains the same. 

The
ground state wave function of the magnetic ion can be written
as \cite{short_pap} $|\Phi \rangle = |t_2^{(1)},t_2^{(2)},t_2^{(3)}\rangle$
and the excited states $|e\rangle = |e^{(1)}\rangle, |e^{(2)}\rangle$
where
\begin{equation}
\begin{array}{rclcl}
t_2^{(1)} \,&=&\,-|d_{3z^2-r^2}\rangle  &+& \eta_1\, 
                  |p_{z}\rangle       \\[3pt] 
t_2^{(2)} \,&=&\,3^{-1/2} [ \sqrt 2 |d_{x^2-y^2}\rangle 
                      +  |d_{zx}\rangle] &+& \eta_2\, 
		         |p_x\rangle \\ [3pt]
t_2^{(3)} \,&=&\,-3^{-1/2} [ \sqrt 2 |d_{xy}\rangle 
                      +  |d_{yz}\rangle ] &+& \eta_2\, 
		         |p_y\rangle \\ [3pt]
  e^{(1)} \,&=&\,3^{-1/2} [ |d_{xy}\rangle 
                      - \sqrt 2 |d_{yz}\rangle ] &+& \eta_3\, 
		        |p_y\rangle  \\ [3pt]
  e^{(2)} \,&=&\,3^{-1/2} [ |d_{x^2-y^2}\rangle 
                      + \sqrt 2 |d_{zx}\rangle] &+& \eta_3\, 
		        |p_x\rangle~~,
\end{array}
\label{states}
\end{equation}
Here the $|d\rangle$ and $|p\rangle$ states are the
Cr-$3d$ and Cr-$4p$ orbitals respectively.
In deriving the above wave functions, we have chosen the axis of
quantization to be parallel to the crystallographic $z$ direction. 
The following convention for the $d$ orbitals is chosen.
\begin{eqnarray*}
d_{xy} & = & -{R(r) \over r^2} \left({15 \over 4\pi}\right)^{{1 \over 2}}
	      xy  \\
d_{yz} & = & {R(r) \over r^2} \left({15 \over 4\pi}\right)^{{1 \over 2}}
	      yz  \\
d_{zx} & = & -{R(r) \over r^2} \left({15 \over 4\pi}\right)^{{1 \over 2}}
	      xz  \\
d_{x^2-y^2} & = & {R(r) \over r^2} \left({15 \over 16\pi}\right)^{{1 \over 2}}
	      (x^2-y^2)  \\
d_{3z^2-r^2} & = & {R(r) \over r^2} \left({5 \over 16\pi}\right)^{{1 \over 2}}
	      (3z^2-r^2)  
\end{eqnarray*}
Here the $R(r)$ are the appropriate radial functions for $3d$ orbitals.
Spin quantum numbers in (\ref{states}) have been suppressed for clarity. 
We also note that 
the expression (\ref{states}) only includes the effect of the hemihedral 
part of the trigonal distortion which is the most dominant interaction. This
interaction is of the form $\eta z$ and leads to a mixing of 
Cr-$3d$ states and Cr-$4p$ states  $(p_x,p_y,p_z)$ 
with coefficients $\eta_1$, $\eta_2$ 
and $\eta_3$ being proportional to $\eta$, the trigonal field. In
addition to this mixing, the spin-orbit coupling also causes a mixing of
the $t_2$ and $e$ orbitals. As explained in our previous paper, it is
this interplay between the trigonal distortion and the spin-orbit
coupling that leads to electric dipole transitions below $T_N$.
Consequently, we focus our attention on those transition matrix
elements in (\ref{jed})
that are proportional to {\em both} the spin orbit interaction
and the trigonal field. It is easy to see that such matrix elements are
of the form
\begin{equation}
\langle t_2|L_zS_z|e\rangle\langle e|{\bf r}|m\rangle\langle
m|E_\mu r_\mu|n\rangle\langle n|E_\nu r_\nu |t_2\rangle ~~,
\label{edmxelem}
\end{equation}
where we have used $\langle m|{\bf p}|n\rangle = -im_e(\omega_n-\omega_m)
\langle m|{\bf r}|n\rangle$, {\em etc.} Let us now consider the case
when the incident light is right circularly polarized \cite{Fiebig_94},
{\it viz.}, ${\bf E} = E \hat{e}_+ $ where $\hat{e}_{\pm}={1 \over
\sqrt{2}} (\hat{e}_x \pm i \hat{e}_y)$. The relevant transition matrix
element is now given by
\[
\sum_{i=2,3\atop j=1,2} \langle t_2^{(i)}|L_zS_z|e^{(j)}\rangle \langle
e^{(j)}|{\bf r}|e^{(j)}\rangle 
\langle e^{(j)}|r_+|n\rangle \langle
n|r_+|t_2^{(i)}\rangle~~,
\]
where the states $|n\rangle$ are higher-energy orbitals of
odd symmetry. Let us now consider that part of the transition matrix
element given above which corresponds to the emission of the second
harmonic, viz.,
$$
\langle t_2^{(i)}|L_zS_z|e^{(j)} \rangle 
\langle e^{(j)}|{\bf r}|e^{(j)} \rangle~.
$$
Since the excited state Cr orbital 
$|e^{(j)}\rangle$ is a mixture of the
Cr-$3d$ and $4p$ states, the emission matrix element has 
contributions of the form
\begin{equation}
\eta \langle t_2^{(i)}|L_zS_z|e^{(j)}\rangle 
\langle p^{(j)}|{\bf r}|e^{(j)}\rangle~,
\label{contribution}
\end{equation}
where $\eta$ is the appropriate admixture of the $p$ 
the $d$ orbitals determined by (\ref{states}). 
Eq.\ \ref{contribution} describes the contribution
 to the ED matrix element of a single
Cr ion. The response of the crystal is given by
the coherent summation over the contributions of
each ion. In Cr$_2$O$_3$ there are two pairs of
sites in the unit cell, the $A/B$ and the $A^\prime/B^\prime$
sites related by inversion symmetry. Since they give
identical contributions we need to consider only one
of the pairs. The contribution of a given unit cell
to the ED matrix element is then proportional to
\begin{equation}
\lambda \left(\eta_A \langle S_z\rangle_A + 
         \eta_B  \langle S_z\rangle_B
        \right)\, 
\label{unit_cell}
\end{equation}
 where $\lambda$ is the spin-orbit coupling constant (see
 ref. \cite{short_pap}) and  $\eta_A$ and $\eta_B$ are
  proportional to the
respective local trigonal fields. Inversion symmetry demands
that $-\eta_B=\eta_A\equiv\eta$. The total contribution
to the ED matrix element is then proportional to
\[
\sum_{\rm all\ unit\ cells}
\lambda\eta (\langle S_z\rangle_A -
            \langle S_z\rangle_B)\, const.
 \equiv \lambda\eta\Delta(T)\,
\]
where $\Delta(T)$ is the antiferromagnetic order
parameter. One can use the explicit form of the orbitals 
given by (\ref{states}) and evaluate the desired matrix elements. For
example, the contribution to SHG by exciting the electron in the
orbital $t_2^{(2)}$ is proportional to
\begin{eqnarray*}
\lambda\eta \triangle(T) & [\langle d_{(x^2-y^2)}|L_z|d_{xy} \rangle
\langle d_{xy}|x|p_y\rangle \times \\
& \langle p_y|z|d_{yz}\rangle \langle d_{yz}|({\bf E}\cdot {\bf r})^2|
d_{zx}\rangle ]~.
\end{eqnarray*}
A similar expression can be written for the contribution arising from 
$t_2^{(3)}$. From our results we find that
(a) the
$t_2^{(1)}$ orbital does not contribute to SHG; This is not surprising
as our mechanism of SHG is mediated through the diagonal
part of the spin-orbit interaction and $L_z t_2^{(1)} = 0$. 
(b) the electric field of the
emitted radiation at frequency $2\omega$ is along the $\hat{x}-i\hat{y}$
direction, i.e., incoming right circularly polarized light generates the
second harmonic with opposite polarization. Likewise, it can be verified
that left circularly polarized light generates a second harmonic that is
right circularly polarized.
Thus our microscopic results reproduce the
selection rules that are given by macroscopic theory. This is because
the expression (\ref{states}) for the Cr$^{3+}$ orbitals has been
obtained by incorporating the full crystal symmetry of Cr$_2$O$_3$.

The constructive interference in emission of the
contributions to the ED matrix elements below $T_N$,
as given by Eq.\ \ref{unit_cell}, is a consequence
of the symmetry of the spin-ordering in Cr$_2$O$_3$. 
We reemphasize that this is because our mechanism 
is a one-ion mechanism that invokes
the {\em diagonal} part of the spin-orbit interaction. 
Photons absorbed by a certain ion are, in this 
mechanism, reemitted by the same ion. The phase
of the emitted electromagnetic wave then depends
on the relative direction of the spin-ordering with
respect to the direction of the trigonal distortion 
at the given site (\ref{unit_cell}). 
The coherent interference of
the waves emitted by all ions may then interfere
constructively. The effect is non-reciprocal since
the direction of the local trigonal distortion
is invariant under time-reversal, whereas the
spin-order is not.  

Let us now consider the case $T > T_N$. Now, inversion is a good
symmetry of the crystal and the electric dipole transitions vanish
identically. In this case, one has to consider the non vanishing term of
the lowest order in the multipole expansion of $\exp (-i{\bf k}_3 \cdot
{\bf r})$. This is, of course, the magnetic dipole term which can be
written as
\begin{eqnarray}
\langle {\bf J}({\bf k}_3,\omega_3) \rangle  =
   {ie^3k_3 \over 4m_e^3 {\hbar}^2}
 q_1q_2 \!\sum_{|m\rangle,|n\rangle}
 \!\langle\Phi|{\bf L}|m\rangle \times
\nonumber \\
\langle m|\exp (-i{\bf k_1}\!\cdot\!{\bf r})a_{1\mu}p_\mu|n\rangle
\langle n|\exp (-i{\bf k_2}\!\cdot\!{\bf r})a_{2\nu}p_\nu|\Phi\rangle 
\nonumber \\
\times  \frac{\exp (i\omega_3t)}{  
(\omega_m-\omega_n+\omega_1) (\omega_n+\omega_2)} + \ldots ~.
\label{jmd}
\end{eqnarray}

The calculation of transition matrix elements in this case is much
simpler as now the dominant contributions contain neither the spin-orbit
interaction nor the trigonal distortion. It is easy to verify that one
only has to consider contributions to the transition matrix elements that
are of the form 
\begin{equation}
\sum_{i=1,2,3} \sum_{j=1,2} \langle t_2^{(i)}|{\bf L}|e^{(j)}\rangle 
\langle
e^{(j)}|r_+|n\rangle\langle n|r_+|t_2^{(i)}\rangle~~,
\label{mdmxelem}
\end{equation}
at any given Cr site. (Here again we have assumed that the incoming
light is right circularly polarized.) The contributions from different
sites can be added up coherently. As in the previous case of the ED
transition, we have also verified using (\ref{states})
that the correct polarization selection rules are reproduced by the
above expression below and above $T_N$.

The microscopic expression for the dynamical current operator together
with the Cr ion wave functions (\ref{states}) enables us to estimate the
magnitudes of the nonlinear susceptibilities. Since we do not
have the correct wave functions of the {\em excited} states, our
estimates are approximate, correct only to an order of magnitude. 
In particular one may also
include contributions from the O-$2p$ orbitals in (\ref{states}).
 Such a contribution would not alter the mechanism for the ED-transition
explained above but would lead to certain quantitative corrections.
From expressions (\ref{source}) and (\ref{jed}),
as also the definition of the macroscopic susceptibility $\chi$ 
we obtain the order of magnitude of $\chi^{(2\omega)}$ as approximately
\begin{eqnarray}
|\chi^{(2\omega)}| ~ \simeq {8n_oe^3  \over \epsilon_o} \left({a_o \over
2\hbar\omega}\right)^2 \times \nonumber \\
{\lambda_o \over a_o} {\hbar \over mc} 
{\lambda \over E_e - E_{t_2}} {\eta \over E_p - E_d} \triangle (T)~~.
\label{chi}
\end{eqnarray}
Here, $\lambda_o\approx 5000\AA$ is the wavelength of the emitted light, 
$a_o\approx 0.69\AA$ is the radius of Cr$^{3+}$, 
$n_o$ is the density of Cr ions in
Cr$_2$O$_3$ ($\simeq 3.3 \times 10^{28}$ m$^{-3}$), 
$\lambda\approx 100$ cm$^{-1}$ 
is the spin-orbit interaction \cite{Rado,ST_58}, $E_{e}-E_{t_2} 
\approx 8000 $cm$^{-1}$ is the difference in energy between the $t_2$ and 
the $e$ orbitals, $\eta\approx 350$ cm$^{-1}$ is the trigonal field 
\cite{ST_58}, $E_{p}-E_{d} \approx 8\times 10^4$ cm$^{-1}$ is the 
difference in energy between the $d$ and the $p$ orbitals that are mixed 
by the trigonal distortion and $\triangle(T)$ is the antiferromagnetic
order parameter. The temperature dependence of $\chi$ is solely through
that of the antiferromagnetic order parameter and consequently $\chi$
vanishes above $T_N$. At temperatures much lower than $T_N$ we find that
$|\chi^{(2\omega)}| \simeq 8.7 \times 10^{-16}$ Coul N$^{-1}$.

Similarly, we find that the magnitude of $\gamma^{(2\omega)}$ is
approximately
\begin{equation}
|\gamma^{(2\omega)}| ~ \simeq {4n_oe^3 \over \epsilon_o}  
\left({a_o \over 2\hbar \omega} \right)^2 {\hbar \over 4mc} ~~.
\label{gamma}
\end{equation}
As explained earlier, this is the MD contribution to SHG 
which depends neither on the trigonal distortion nor the spin-orbit
interaction. It is also independent of temperature and we estimate it to
be of the order of $11 \times 10^{-16}$ Coul N$^{-1}$.
The contributions to
${\bf J}$ from the nonlinear susceptibilities $\chi$ and $\gamma$ can
produce interference effects in SHG only if the first two terms in the
r.h.s. of equation (\ref{source}) are of the same order of magnitude, 
{\it viz.}, $|\chi| \simeq |\gamma|$.  From our estimates of $\chi$ and 
$\gamma$ above we find that this is indeed satisfied thereby leading to an
interference effect in SHG which is non reciprocal and which vanishes
above $T_N$.

\section{Gyrotropic Birefringence and the Optical Magnetoelectric
Effect}
The ED transition in the optical region that is allowed below $T_N$ in
Cr$_2$O$_3$ can also be seen in one-photon experiments.  In this
section, we consider the phenomena of Gyrotropic Birefringence (GB) and
the associated Optical Magnetoelectric Effect that are one-photon
processes in contrast to SHG which is a two-photon process. 

GB is a non-reciprocal optical effect that appears as a shift in the
principal optical axis along with a change in the velocity of
propagation of light. The possibility of observing this effect in
Cr$_2$O$_3$ and using it to distinguish between antiferromagnetic
domains was first pointed out by Brown and co-workers \cite{Brown_63}.
Their analysis of the problem was purely macroscopic and they showed
that GB appears in the form of a polar $c$ tensor (one that changes sign
under both space and time inversion) below $T_N$. Later, Hornreich and
Shtrikman \cite{HS_68} presented the first quantum mechanical treatment
of this problem. They showed that the gyrotropic birefringence tensor
can be described in terms of electric quadrupole and magnetoelectric
effects. From their results, they estimated that at optical frequencies,
the induced rotation of the principal optical axes of Cr$_2$O$_3$ would
be of the order of $10^{-8} - 10^{-6}$ rad and the magnetoelectric
susceptibility would be of the order of $10^{-8}$ (in dimensionless
units). This phenomenon was also analyzed by Graham and Raab \cite{gra}
using a multipole theory of wave propagation.

Recently however, Krichevtsov {\it et al.} \cite{Krichevtsov_93}
observed for the first time,
spontaneous non-reciprocal rotation of the optical axes in Cr$_2$O$_3$.
They found that the observed rotation and the magnetoelectric
susceptibility were 4 orders of magnitude larger than those predicted by
Hornreich and Shtrikman. They also found that the observed temperature
dependence of these non reciprocal effects corresponds roughly to that
of the antiferromagnetic order parameter, something that does not follow
obviously from previous calculations \cite{HS_68}.
The intensity of these effects
and the temperature dependence led them to speculate that these effects
were attributable to ED transitions in the optical range. We now show that
this is indeed the case and the mechanism we discussed earlier does lead
to effects that are of the same order of magnitude as those observed. 

To see this, we calculate the magnetoelectric susceptibility in the
optical region, allowing for an ED transition below $T_N$. The
calculation is in many ways similar to that presented earlier for the
nonlinear susceptibilities. The magnetoelectric susceptibility is
defined as ${\bf M}^{(\omega)} = \alpha : {\bf E}^{(\omega)}$ and we
estimate it by calculating directly, the quantity 
$\langle {\bf M} \rangle = g\mu_B \langle {\bf L} +2{\bf S} \rangle$
in perturbation theory. Since in this case we are only interested in
temperatures below the N\'eel temperature, we use the electric dipole
approximation and consider the linear response of the system. It is easy
to verify that the induced magnetization is given by
\begin{eqnarray*}
\langle {\bf M} \rangle & = & {e\over\hbar}{\rm Re} \Big[
\sum_{|m\rangle}
\langle \Phi |{\bf M}|m\rangle{\langle m|{\bf E}\cdot{\bf r}|\Phi
 \rangle \over
(\omega - \omega_m)}\exp(-i\omega_t) \\
& & +\,\langle \Phi |{\bf M}|m\rangle{\langle m|{\bf E}\cdot{\bf r}|\Phi
\rangle 
\over (\omega + \omega_m)}\exp(+i\omega_t) \Big] ~~,
\end{eqnarray*}
where we have assumed the electric field to be of the form ${\bf
E} = {\bf E}_o \cos ({\bf k}\cdot {\bf r} -\omega t)$.
Choosing (as in the experiment
\cite{Krichevtsov_93}) the incident light along the $ \hat{x}$ direction
and the electric field ${\bf E} \| \hat{z}$ , it is easy to see that a
typical matrix element that would contribute to the magnetoelectric
susceptibility is of the form,
$$
\langle t_2| L_z |e\rangle \langle e | L_zS_z |t_2\rangle \langle t_2|
zE_z |t_2\rangle~~.
$$
Note that $\langle t_2| zE_z |t_2\rangle$ would be proportional to the
trigonal mixing of the $3d$ and the $4p$ orbitals.
A straightforward calculation using
(\ref{states}) to evaluate appropriate transition matrix elements gives
us the expression for the magnetoelectric susceptibility,
\begin{eqnarray}
\alpha_{xx}  \sim ~ 4 \mu_o c e~{g\mu_B \over \hbar (\omega-\omega_m)} \times 
\nonumber \\
{\lambda\over E_e-E_{t_2}}\ {\eta\over E_p-E_d}~n_o~ \triangle(T)~,
\end{eqnarray}
in dimensionless units. Here, $n_o$ is the density of Cr ions in
Cr$_2$O$_3$ ($\simeq 3.3 \times 10^{28}$ m$^{-3}$) and
 $\hbar(\omega-\omega_n) \sim 0.5$ eV in the region of experimental
 interest.  Thus by evaluating the above expression, we
estimate $\alpha_{xx} \sim 0.2 \times 10^{-4}$ which is of the same
order of magnitude as that observed experimentally. This also means that the
non-reciprocal rotation would be $\sim 10^{-4}$ rad. Since the ED
process that we consider couples light to the order parameter, the observed 
temperature dependence follows naturally from our theory.

\section{Summary and Conclusions}
In this paper we have developed a microscopic theory of
non-reciprocal optical effects observed below $T_N$ in Cr$_2$O$_3$. 
We have shown that these effects can be explained by an electric dipole 
process that arises from an interplay between the spin-orbit coupling 
and the trigonal distortion of the ligand field. Such a process 
couples light directly to the antiferromagnetic order parameter.
In contrast to other ED mechanisms that have been considered in
the literature so far \cite{several}, 
our mechanism is a one-ion mechanism that couples
light to the sublattice magnetization rather than any magnetic excitations.
Photons absorbed by a certain Cr$^{3+}$ ion are also emitted by
the same ion. The coupling to the antiferromagnetic order 
parameter then occurs throught constructive interference of the
photons emitted by different Cr-ions.

It might seem that our mechanism leads to effects that are weak
as they involve the trigonal distortion and spin-orbit
interaction, both weak effects by themselves. However their interplay
leads to an electric dipole transition whose oscillator strength is
large in the optical region. Consequently, observed effects are strong.
Though we have used this mechanism to explain successfully the 
phenomena of Second Harmonic Generation, Gyrotropic Birefringence and 
the Optical Magnetoelectric effect that have been observed experimentally in
Cr$_2$O$_3$, our theory can be generalized to all materials where
i) the magnetic ion is not at a center of inversion and ii) 
inversion is still a macroscopic symmetry above $T_N$ 
but is broken below $T_N$ due to the ordering of the magnetic ions. 
Thus, non-reciprocal effects should be observable, for example, in  
the cuprate Gd$_2$CuO$_4$ 
below the ordering temperature of the Gadolinium magnetic subsystem, 
$T_N$(Gd)=6.5K \cite{GdCuO} when inversion symmetry is broken as also in
V$_2$O$_3$ and MnTiO$_3$.

\section{Appendix}

   The components for the nonlinear susceptibilities ($\chi$) and 
($\gamma$) in Table I are as follows: 

\begin{eqnarray}
\chi_e&\equiv&\chi_{yyy}=-\chi_{yxx}=-\chi_{xyx}=-\chi_{xxy} \nonumber \\
\chi_{e_1}&\equiv&\chi_{yyy} \nonumber\\
\chi_{e_2}&\equiv&-\chi_{yxx}=-\chi_{xyx}=-\chi_{xxy}
\nonumber  \\
\chi_a&\equiv&\chi_{yxz}=-\chi_{xyz} \nonumber  \\
\chi_{a_1}&\equiv&\chi_{yxz} \nonumber \\
\chi_{a_2}&\equiv&-\chi_{xyz} \nonumber \\
\chi_b&\equiv&\chi_{yzx}=-\chi_{xzy} \nonumber  \\
\chi_{b_1}&\equiv&\chi_{yzx}  \nonumber  \\
\chi_{b_2}&\equiv&-\chi_{xzy}  \nonumber \\
\chi_1&\equiv&\chi_{zzy}=\chi_{zyz}=\chi_{yzz}  \nonumber  \\
\chi_2&\equiv&\chi_{zyx}  \nonumber \\
\chi_3&\equiv&\chi_{xxx}  \nonumber  \\
\chi_4&\equiv&\chi_{xyy}=\chi_{yyx}=\chi_{yxy}  \nonumber  \\
\chi_5&\equiv&\chi_{zxx}=\chi_{xxz}=\chi_{xzx} \nonumber  \\
\chi_6&\equiv&\chi_{zyy}=\chi_{yzy}=\chi_{yyz} \nonumber \\
\chi_7&\equiv&\chi_{zzx}=\chi_{zxz}=\chi_{xzz}  \nonumber  \\
\chi_8&\equiv&\chi_{zzz}  \nonumber \\
\chi_9&\equiv&\chi_{yyz}=\chi_{yzy}=\chi_{zyy} \nonumber  \\
\chi_{10}&\equiv&\chi_{zxy}  \nonumber  \\
\gamma_m&\equiv&\gamma_{yyy}=-\gamma_{yxx}=-\gamma_{xyx}=-\gamma_{xxy}
  \nonumber \\
\gamma_{m_1}&\equiv&\gamma_{yyy}   \nonumber \\
\gamma_{m_2}&\equiv&-\gamma_{xxy}=-\gamma_{xyx}=-\gamma_{yxx}  \nonumber  \\
\gamma_1&\equiv&\gamma_{zzy}=\gamma_{zyz}=\gamma_{yzz}   \nonumber
\end{eqnarray}
Above the N\'eel temperature all $\chi$'s and $\gamma_1$
vanish (and $\gamma_{m_1}=\gamma_{m_2}\equiv\gamma_m$).
It is then reasonable to assume that below $T_N$ $\gamma_1$ is
much smaller than all other matrix elements since it
is of magnetic dipole character and only allowed due to
the breaking of inversion symmetry below the N\'eel
temperature. The components of the electric quadrupole susceptibilities
 in Table II
are as follows.
\begin{eqnarray}
\tilde{\chi}_1 & \equiv & \tilde{\chi}_{xxyy}= \tilde{\chi}_{yyxx} \nonumber\\
\tilde{\chi}_2 & \equiv & \tilde{\chi}_{xxzz}= \tilde{\chi}_{yyzz} \nonumber\\
\tilde{\chi}_3 & \equiv & \tilde{\chi}_{zzxx}= \tilde{\chi}_{zzyy} \nonumber\\
\tilde{\chi}_4 & \equiv & -\tilde{\chi}_{zxyy}= \tilde{\chi}_{zxxx} =
-\tilde{\chi}_{zyxy} \nonumber\\
& = & -\tilde{\chi}_{zyyx} = -\tilde{\chi}_{xyzy}= -\tilde{\chi}_{xyyz} = 
-\tilde{\chi}_{xzyy} \nonumber
\end{eqnarray}

\acknowledgements
We acknowledge several discussions with T. Djelani, M. Fiebig,
D. Fr\"ohlich, G. Sluy\-ter\-man v. L. and H. J. Thiele. 
This work was 
supported by the Deutsche Forschungsgemeinschaft, the Graduiertenkolleg 
``Festk\"{o}rperspektroskopie'' and by the European Community Human Capital 
and Mobility program.

%
%
%
\newpage\phantom{just to fill the page}{\tiny .}
\newpage
\widetext
%
%
\[
\begin{array}{r|c|c|c|}
&\multicolumn{3}{c|}{direction\ of\ incident\ light}\\
\cline{2-4}
& z & y & x \\
\hline 
& \phantom{,}& & \\
z\ &
\left(\begin{array}{c} \gamma_m (E_x^2-E_y^2)-2\chi_e E_xE_y\\
                       -\chi_e(E_x^2-E_y^2)-2\gamma_mE_xE_y\\
                       0
\end{array}\right)
&
\left(\begin{array}{c} 0\\
                       -(\chi_a+\chi_b)E_xE_z-\chi_eE_xE_x\\
                       0
\end{array}\right)
&
\left(\begin{array}{c} (\chi_a+\chi_b)E_yE_z\\
                       \chi_{e} E_y^2\\
                       \gamma_{m} E_y^2
\end{array}\right) \\
& \phantom{,}& & \\
\hline
& \phantom{,}& & \\
y\ &
\left(\begin{array}{c} (\chi_3+\gamma_{m_2})E_x^2+(\chi_4-\gamma_{m_1})E_y^2\\
                       2(\chi_4-\gamma_{m_2})E_xE_y\\
                       \chi_5E_x^2+\chi_6E_y^2
\end{array}\right)
&
\left(\begin{array}{c} \chi_3E_x^2+\chi_7E_z^2+2\chi_5E_xE_z\\
                       0\\
                       \chi_5E_x^2+\chi_8E_z^2+2\chi_7E_xE_z
\end{array}\right)
&
\left(\begin{array}{c} \chi_4E_y^2+\chi_7E_z^2\\
                       2(\chi_9-\gamma_1)E_yE_z\\
                       (\chi_9+\gamma_{m_1}) E_y^2+(\chi_8+\gamma_1)E_z^2
\end{array}\right) \\
& \phantom{,}& & \\
\hline
& \phantom{,}& & \\
x\ &
\left(\begin{array}{c} \gamma_{m_2}E_x^2-\gamma_{m_1}E_y^2-2\chi_{e_2}E_xE_y\\
                       -\chi_{e_2}E_x^2+\chi_{e_1}E_y^2-2\gamma_{m_2}E_xE_y\\
                       (\chi_{10}+\chi_{2})E_xE_y
\end{array}\right)
&
\left(\begin{array}{c} 0\\
                       -\chi_{e_2}E_x^2+\chi_{1}E_z^2+(\chi_{a_1}+
                       \chi_{b_1})E_xE_z\\
                       0
\end{array}\right)
&
\left(\begin{array}{c} -(\chi_{a_2}+\chi_{b_2})E_yE_z\\
                       \chi_{e_1}E_y^2+\chi_{1}E_z^2-2\gamma_1E_yE_z\\
                       \gamma_{m_1}E_y^2+\gamma_1E_z^2+2\chi_{1}E_yE_z
\end{array}\right) \\
& \phantom{,}& & \\
\hline
\end{array}
\]
%

\begin{table}
\caption{Predictions for the componentsof the source term $\bf S$ from 
macroscopic symmetry considerations. A constant term $4 \omega^2/ c^2$
has here been omitted for simplicity.  The three columns are
the predictions for light incident in $z$, $y$ and $x$ directions
respectively. The three rows are for the moment of the
Cr$^{3+}$ spins aligned parallel to the crystallographic
$z$, $y$ and $x$ axis respectively. $\gamma_i$ ($i=m,m_1,m_2,1$) 
and $\chi_i$ 
 ($i=e,e_1,e_2,a,a_1,a_2,b,b_1,b_2,1,2,3,...,10$)
correspond to the remaining independent components for the 
magnetic and electric susceptibilities respectively. 
See the appendix for an
explicit account of these components.}
\end{table}

%
\[
\begin{array}{r|c|c|c|}
&\multicolumn{3}{c|}{direction\ of\ incident\ light}\\
\cline{2-4}
& z & y & x \\
\hline 
& \phantom{,}& & \\
z\ &
\left(\begin{array}{c} \tilde{\chi}_4 (E_x^2-E_y^2)\\
                       -2 \tilde{\chi}_4E_xE_y\\
                       \tilde{\chi}_3 (E_x^2+E_y^2)
\end{array}\right)
&
\left(\begin{array}{c} 0\\
                       \tilde{\chi}_1 E_x^2 + \tilde{\chi}_2 E_z^2\\
                       0
\end{array}\right)
&
\left(\begin{array}{c} \tilde{\chi}_2 E_z^2 + \tilde{\chi}_1 E_y^2\\
                       - 2 \tilde{\chi}_4 E_yE_z\\
                       -\tilde{\chi}_4 E_y^2
\end{array}\right) \\
& \phantom{,}& & \\
\hline
\end{array}
\]

\begin{table}
\caption{Electric quadrupole contributions to the source term, $\bf S$.
 The constant $4 \omega^2/ c^2$ has been omitted for simplicity.
 The three columns are the predictions for light incident in
 $z$, $y$ and $x$ directions respectively and for the 
  Cr$^{3+}$ spins aligned parallel to the $z$ axis.
 $\tilde{\chi}_i$ ($i=1,2,3,4$) correspond to the remaining independent
 components.  (See the appendix). }
\end{table}
 
\end{document}